\documentclass[runningheads]{svmult}

\usepackage{makeidx}   % allows index generation
\usepackage{graphicx}  % standard LaTeX graphics tool
\usepackage{subeqnar}  % subnumbers individual equations
\usepackage{multicol}  % used for the two-column index
\usepackage{physprbb}  % modified textarea for proceedings,
\makeindex             % used for the subject index
\begin{document}
\title*{The Environmental Impact of Supermassive Black Holes}
\toctitle{The Environmental Impact of Supermassive Black Holes}
\titlerunning{The Environmental Impact of Supermassive Black Holes}
\author{Abraham Loeb}
\authorrunning{Avi Loeb}
\institute{Astronomy Department, Harvard University, Cambridge MA 02138,
USA}

\maketitle % typesets the title of the contribution

\begin{abstract}
The supermassive black holes observed at the centers of almost all
present-day galaxies, had a profound impact on their environment. I
highlight the principle of {\it self-regulation}, by which supermassive
black holes grow until they release sufficient energy to unbind the gas
that feeds them from their host galaxy. This principle explains several
observed facts, including the correlation between the mass of a central
black hole and the depth of the gravitational potential well of its host
galaxy, and the abundance and clustering properties of bright quasars in
the redshift interval of $z\sim 2$--6.  At lower redshifts, quasars might
have limited the maximum mass of galaxies through the suppression of
cooling flows in X-ray clusters.  The seeds of supermassive black holes
were likely planted in dwarf galaxies at redshifts $z> 10$, through the
collapse of massive or supermassive stars.  The minimum seed mass can be
identified observationally through the detection of gravitational waves
from black hole binaries by {\it Advanced LIGO} or {\it LISA}.  Aside from
shaping their host galaxies, quasar outflows filled the intergalactic
medium with magnetic fields and heavy elements.  Beyond the reach of these
outflows, the brightest quasars at $z>6$ have ionized exceedingly large
volumes of gas (tens of comoving Mpc) prior to global reionization, and
must have suppressed the faint end of the galaxy luminosity function in
these volumes before the same occurred through the rest of the universe.

\end{abstract}

\section{The Principle of Self-Regulation}

The fossil record in the present-day universe indicates that every bulged
galaxy hosts a supermassive black hole (BH) at its center
\cite{Kor03}. These BHs are dormant or faint most of the time, but
ocassionally flash in a short burst of radiation that lasts for a small
fraction of the Hubble time. The short duty cycle acounts for the fact that
bright quasars are much less abundant than their host galaxies, but it begs
the more fundamental question: {\it why is the quasar activity so brief?}
A natural explanation is that quasars are suicidal, namely the energy
output from the BHs regulates their own growth.

Supermassive BHs make up a small fraction, $< 10^{-3}$, of the total mass
in their host galaxies, and so their direct dynamical impact is limited to
the central star distribution where their gravitational influence
dominates. Dynamical friction on the background stars keeps the BH close to
the center. Random fluctuations in the distribution of stars induces a
Brownian motion of the BH. This motion can be decribed by the same Langevin
equation that captures the motion of a massive dust particle as it responds
to random kicks from the much lighter molecules of air around it
\cite{Cha02}.  The characteristic speed by which the BH wanders around the
center is small, $\sim (m_\star/M_{\rm BH})^{1/2}\sigma_\star$, where
$m_\star$ and $M_{\rm BH}$ are the masses of a single star and the BH,
respectively, and $\sigma_\star$ is the stellar velocity dispersion. Since
the random force fluctuates on a dynamical time, the BH wanders across a
region that is smaller by a factor of $\sim (m_\star/M_{\rm BH})^{1/2}$
than the region traversed by the stars inducing the fluctuating force on
it.

The dynamical insignificance of the BH on the global galactic scale is
misleading. The gravitational binding energy per rest-mass energy of
galaxies is of order $\sim (\sigma_\star/c)^2< 10^{-6}$.  Since BH are
relativistic objects, the gravitational binding energy of material that
feeds them amounts to a substantial fraction its rest mass energy. Even if
the BH mass occupies a fraction as small as $\sim 10^{-4}$ of the baryonic
mass in a galaxy, and only a percent of the accreted rest-mass energy leaks
into the gaseous environment of the BH, this slight leakage can unbind the
entire gas reservoir of the host galaxy! This order-of-magnitude estimate
explains why quasars are short lived.  As soon as the central BH accretes
large quantities of gas so as to significantly increase its mass, it
releases large amounts of energy that would suppress further accretion onto
it. In short, the BH growth is {\it self-regulated}.

The principle of {\it self-regulation} naturally leads to a correlation
between the final BH mass, $M_{\rm bh}$, and the depth of the gravitational
potential well to which the surrounding gas is confined, $\sim
\sigma_\star^2$. Indeed such a correlation is observed in the present-day
universe \cite{Tre02}. The observed power-law relation between $M_{\rm bh}$
and $\sigma_\star$ can be generalized to a correlation between the BH mass
and the circular velocity of the host halo, $v_c$ \cite{Fer02}, which in
turn can be related to the halo mass, $M_{\rm halo}$, and redshift, $z$
\cite{WL03}
\begin{eqnarray}
\label{eq:1}
\nonumber M_{\rm bh}(M_{\rm halo},z) &=&\mbox{const} \times v_c^5\\
&&\hspace{-25mm}= \epsilon_{\rm o} M_{\rm halo} \left(\frac{M_{\rm
halo}}{10^{12}M_{\odot}}\right)^{\frac{2}{3}}
[\zeta(z)]^\frac{5}{6}(1+z)^\frac{5}{2},
\end{eqnarray}
where $\epsilon_{\rm o}\approx 10^{-5.7}$ is a constant, $\zeta(z)$ is
close to unity and defined as $\zeta\equiv
[(\Omega_m/\Omega_m^z)(\Delta_c/18\pi^2)]$, $\Omega_m^z \equiv
[1+(\Omega_\Lambda/\Omega_m)(1+z)^{-3}]^{-1}$,
$\Delta_c=18\pi^2+82d-39d^2$, and $d=\Omega_m^z-1$ (see equations~22--25 in
Ref. \cite{Bar01} for the relation between $v_c$ and $M_{\rm halo}$).  If
quasars shine near their Eddington limit as suggested by observations of
low and high-redshift quasars \cite{Flo03,Wil03}, then the above value of
$\epsilon_{\rm o}$ implies that a fraction of $\sim 5$--$10\%$ of the
energy released by the quasar over a galactic dynamical time needs to be
captured in the surrounding galactic gas in order for the BH growth to be
self-regulated \cite{WL03}.

\begin{figure}[ht]
\begin{center}
\includegraphics[width=.95\textwidth]{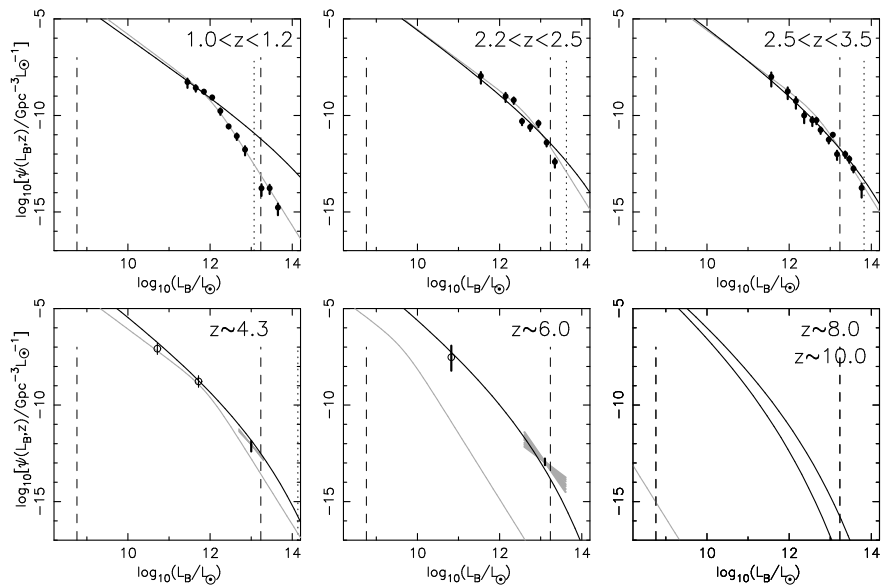}
\caption[]{Comparison of the observed and model luminosity functions (from
\cite{WL03}). The data points at $z<4$ are summarized in Ref. \cite{Pei95},
while the light lines show the double power-law fit to the {\it 2dF} quasar
luminosity function \cite{Boy00}.  At $z\sim4.3$ and $z\sim6.0$ the data is
from Refs. \cite{Fan01a}. The grey regions show the 1-$\sigma$ range of
logarithmic slope ($[-2.25,-3.75]$ at $z\sim4.3$ and $[-1.6,-3.1]$ at
$z\sim6$), and the vertical bars show the uncertainty in the
normalization. The open circles show data points converted from the X-ray
luminosity function \cite{Bar03} of low luminosity quasars using the median
quasar spectral energy distribution.  In each panel the vertical dashed
lines correspond to the Eddington luminosities of BHs bracketing the
observed range of the $M_{\rm bh}$--$v_{\rm c}$ relation, and the vertical
dotted line corresponds to a BH in a $10^{13.5}M_\odot$ galaxy.}
\label{fig1}
\end{center}
\end{figure}

With this interpretation, the $M_{\rm bh}$--$\sigma_\star$ relation
reflects the limit introduced to the BH mass by self-regulation; deviations
from this relation are inevitable during episodes of BH growth or as a
result of mergers of galaxies that have no cold gas in them.  A physical
scatter around this upper envelope could also result from variations in the
efficiency by which the released BH energy couples to the surrounding gas.

\begin{figure}[h]
\begin{center}
\includegraphics[width=.7\textwidth]{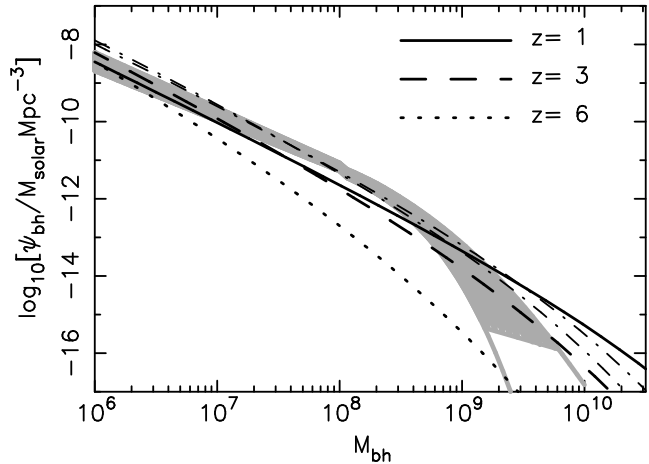}
\end{center}
\caption[]{The comoving density of supermassive BHs per unit BH mass (from
\cite{WL03}). The grey region shows the estimate based on the observed
velocity distribution function of galaxies in Ref. \cite{She03} and the
$M_{\rm bh}$--$v_c$ relation in Eq. (\ref{eq:1}). The lower bound
corresponds to the lower limit in density for the observed velocity
function while the grey lines show the extrapolation to lower densities. We
also show the mass function computed at $z=1$, 3 and 6 from the
Press-Schechter\cite{PS74} halo mass function and Eq.~(\ref{eq:1}), as well
as the mass function at $z\sim2.35$ and $z\sim3$ implied by the observed
density of quasars and a quasar lifetime of order the dynamical time of the
host galactic disk, $t_{\rm dyn}$ (dot-dashed lines).}
\label{fig2}
\end{figure}

\begin{figure}[ht]
\begin{center}
\includegraphics[width=.8\textwidth]{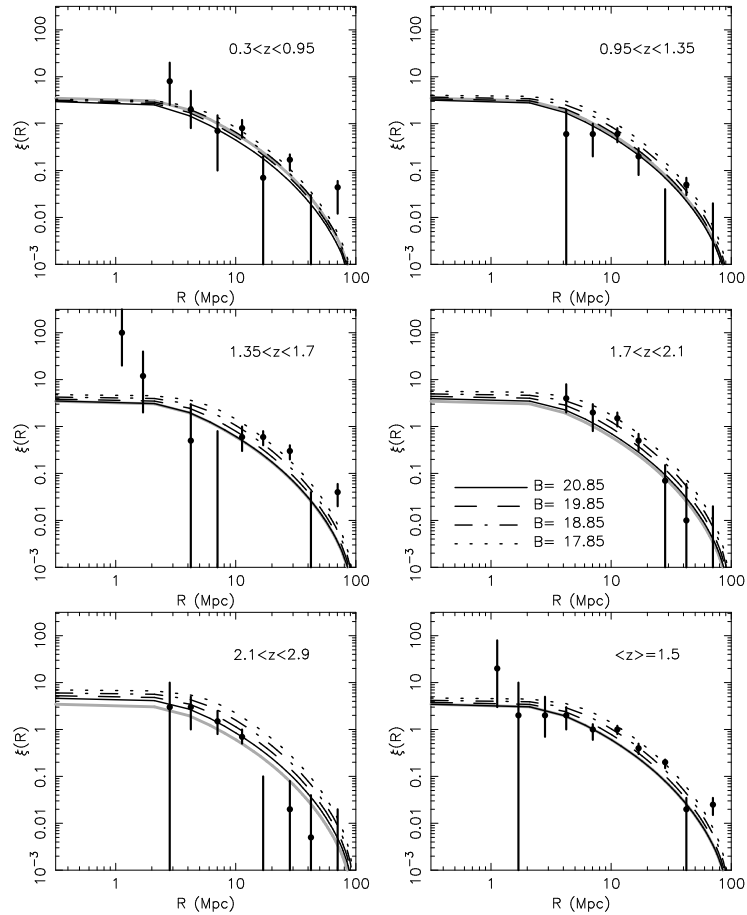}
\end{center}
\caption[]{Predicted correlation function of quasars at various redshifts
in comparison to the {\it 2dF} data \cite{Cro01} (from \cite{WLcl}). The
dark lines show the correlation function predictions for quasars of various
apparent B-band magnitudes. The {\it 2dF} limit is $B\sim20.85$. The lower
right panel shows data from entire {\it 2dF} sample in comparison to the
theoretical prediction at the mean quasar redshift of $\langle
z\rangle=1.5$. The $B=20.85$ prediction at this redshift is also shown by
thick gray lines in the other panels to guide the eye. The predictions are
based on the scaling $M_{\rm bh}\propto v_c^5$ in Eq. (\ref{eq:1}).  }
\label{fig3}
\end{figure}

Various prescriptions for self-regulation were sketched by Silk \& Rees
\cite{Sil98}. These involve either energy or momentum-driven winds, where
the latter type is a factor of $\sim v_c/c$ ~less efficient
\cite{Beg04,Kin03,Mur04}. Wyithe \& Loeb \cite{WL03} demonstrated that a
particularly simple prescription for an energy-driven wind can reproduce
the luminosity function of quasars out to highest measured redshift, $z\sim
6$ (see Figs. \ref{fig1} and \ref{fig2}), as well as the observed
clustering properties of quasars at $z\sim 3$ \cite{WLcl} (see
Fig. \ref{fig3}). The prescription postulates that: {\it (i)}
self-regulation leads to the growth of $M_{\rm bh}$ up the
redshift-independent limit as a function of $v_c$ in Eq. (\ref{eq:1}), for
all galaxies throughout their evolution; and {\it (ii)} the growth of
$M_{\rm bh}$ to the limiting mass in Eq. (\ref{eq:1}) occurs through halo
merger episodes during which the BH shines at its Eddington luminosity
(with the median quasar spectrum) over the dynamical time of its host
galaxy, $t_{\rm dyn}$.  This model has only one adjustable parameter,
namely the fraction of the released quasar energy that couples to the
surrounding gas in the host galaxy. This parameter can be fixed based on
the $M_{\rm bh}$--$\sigma_\star$ relation in the local universe
\cite{Fer02}.  It is remarkable that the combination of the above simple
prescription and the standard $\Lambda$CDM cosmology for the evolution and
merger rate of galaxy halos, lead to a satisfactory agreement with the rich
data set on quasar evolution over cosmic history.

The cooling time of the heated gas is typically longer than its dynamical
time and so the gas should expand into the galactic halo and escape the
galaxy if its initial temperature exceeds the virial temperature of the
galaxy \cite{WL03}. The quasar remains active during the dynamical time of
the initial gas reservoir, $\sim 10^7$ years, and fades afterwards due to
the dilution of this reservoir.  Accretion is halted as soon as the quasar
supplies the galactic gas with more than its binding energy. The BH growth
may resume if the cold gas reservoir is replenished through a new merger.

Agreement between the predicted and observed correlation function of
quasars (Fig. \ref{fig3}) is obtained only if the BH mass scales with
redshift as in Eq. (\ref{eq:1}) and the quasar lifetime is of the
order of the dynamical time of the host galactic disk \cite{WLcl},
\begin{equation}
t_{\rm dyn}= 10^7~[\xi(z)]^{-1/2}\left({1+z\over 3}\right)^{-3/2}~{\rm yr}.
\label{eq:life}
\end{equation}

The inflow of cold gas towards galaxy centers during the growth phase of
the BH would naturally be accompanied by a burst of star formation.  The
fraction of gas that is not consumed by stars or ejected by supernovae,
will continue to feed the BH. It is therefore not surprising that quasar
and starburst activities co-exist in Ultra Luminous Infrared Galaxies
\cite{Gen02}, and that all quasars show broad metal lines indicating a
super-solar metallicity of the surrounding gas \cite{Ham03}. Applying a
similar self-regulation principle to the stars, leads to the expectation
\cite{WL03,Kau00} that the ratio between the mass of the BH and the mass in
stars is independent of halo mass (as observed locally \cite{Mag98}) but
increases with redshift as $\propto \xi(z)^{1/2}(1+z)^{3/2}$. A
consistent trend has indeed been inferred in an observed sample of
gravitationally-lensed quasars \cite{Rix99}.

The upper mass of galaxies may also be regulated by the energy output from
quasar activity. This would account for the fact that cooling flows are
suppressed in present-day X-ray clusters \cite{Fab04,Boe02,OhS04}, and that
massive BHs and stars in galactic bulges were already formed at $z\sim
2$. The quasars discovered by the {\it Sloan Digital Sky Survey} ({\it
SDSS}) at $z\sim 6$ mark the early growth of the most massive BHs and
galactic spheroids. The present-day abundance of galaxies capable of
hosting BHs of mass $\sim 10^9M_\odot$ (based on Eq. \ref{eq:1}) already
existed at $z\sim 6$ ~\cite{Loe03}. At some epoch, the quasar energy output
may have led to the extinction of cold gas in these galaxies and the
suppression of further star formation in them, leading to an apparent
``anti-hierarchical'' mode of galaxy formation where massive spheroids
formed early and did not make new stars at late times. In the course of
subsequent merger events, the cores of the most massive spheroids acquired
an envelope of collisionless matter in the form of already-formed stars or
dark matter \cite{Loe03}, without the proportional accretion of cold gas
into the central BH. The upper limit on the mass of the central BH and the
mass of the spheroid is caused by the lack of cold gas and cooling flows in
their X-ray halos. In the cores of cooling X-ray clusters, there is often
an active central BH that supplies sufficient energy to compensate for the
cooling of the gas \cite{Boe02,Fab04,Beg04}. The primary physical process
by which this energy couples to the gas is still unknown.

\begin{figure}[ht]
\begin{center}
\includegraphics[width=.95\textwidth]{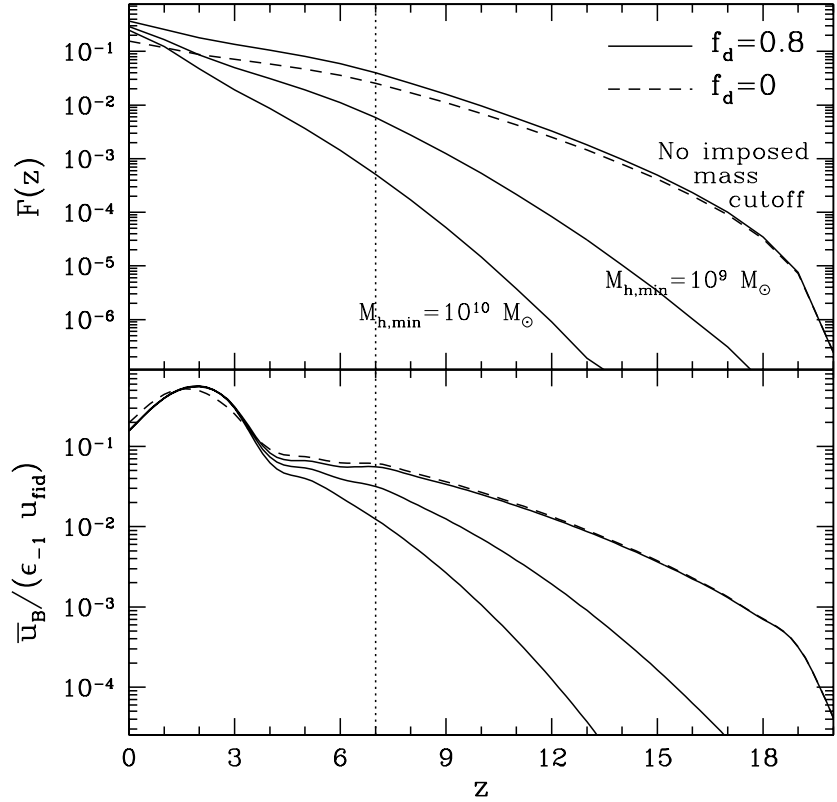}
\end{center}
\caption[]{ The global influence of magnetized quasar outflows on the
intergalactic medium (from \cite{Fur01}). {\it Upper Panel:} Predicted
volume filling fraction of magnetized quasar bubbles $F(z)$, as a function
of redshift.  \emph{Lower Panel:} Ratio of normalized magnetic energy
density, $\bar{u}_B/\epsilon_{-1}$, to the fiducial thermal energy density
of the intergalactic medium $u_{fid} = 3 n(z) k T_{IGM}$, where $T_{IGM} =
10^4 ~{\rm K}$, as a function of redshift (see \cite{Fur01} for more
details).  In each panel, the solid curves assume that the blast wave
created by quasar ouflows is nearly (80\%) adiabatic, and that the minimum
halo mass of galaxies, $M_{h,min}$, is determined by atomic cooling before
reionization and by suppression due to galactic infall afterwards (top
curve), $M_{h,min} = 10^9 M_\odot$ (middle curve), and $M_{h,min} = 10^{10}
M_\odot$ (bottom curve). The dashed curve assumes a fully-radiative blast
wave and fixes $M_{h,min}$ by the thresholds for atomic cooling and infall
suppression.  The vertical dotted line indicates the assumed redshift of
complete reionization, $z_r=7$.  }
\label{fig4}
\end{figure}

\section{Feedback on Large Intergalactic Scales}

Aside from affecting their host galaxy, quasars disturb their large-scale
cosmological environment. Powerful quasar outflows are observed in the form
of radio jets \cite{Beg84} or broad-absorption-line winds
\cite{Bran03}. The amount of energy carried by these outflows is largely
unknown, but could be comparable to the radiative output from the same
quasars. Furlanetto \& Loeb \cite{Fur01} have calculated the intergalactic
volume filled by such outflows as a function of cosmic time (see
Fig. \ref{fig4}). This volume is likely to contain magnetic fields and
metals, providing a natural source for the observed magnetization of the
metal-rich gas in X-ray clusters \cite{Kro01} and in galaxies \cite{Dal90}.
The injection of energy by quasar outflows may also explain the deficit of
Ly$\alpha$ absorption in the vicinity of Lyman-break galaxies
\cite{Ade03,Cro02} and the required pre-heating in X-ray clusters
\cite{Bor02,Boe02}.

\begin{figure}[ht]
\begin{center}
\includegraphics[width=.8\textwidth]{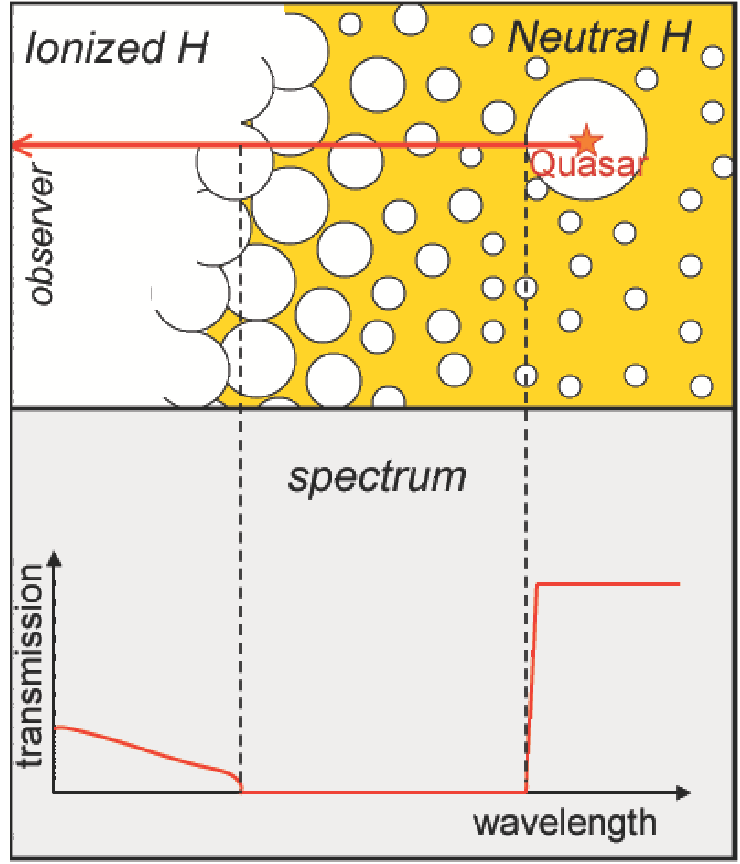}
\end{center}
\caption[]{Quasars serve as probes of the end of reionization. The measured
size of the HII regions around {\it SDSS} quasars can be used
\cite{WL04b,Mes04} to demonstrate that a significant fraction of the
intergalactic hydrogen was neutral at $z\sim 6.3$ or else the inferred size
of the quasar HII regions would have been much larger than observed
(assuming typical quasar lifetimes \cite{Mar03}).  Also, quasars can be
used to measure the redshift at which the intergalactic medium started to
transmit Ly$\alpha$ photons\cite{Whi03,WL04c}. The upper panel illustrates how
the line-of-sight towards a quasar intersects this transition redshift. The
resulting Ly$\alpha$ transmission of the intrinsic quasar spectrum is shown
schematically in the lower panel.}
\label{fig5}
\end{figure}

Beyond the reach of their outflows, the brightest {\it SDSS} quasars at
$z>6$ are inferred to have ionized exceedingly large regions of gas (tens
of comoving Mpc) around them prior to global reionization (see
Fig. \ref{fig5} and Refs. \cite{Whi03,WL04c}). Thus, quasars must have
suppressed the faint-end of the galaxy luminosity function in these regions
before the same occurred throughout the universe.  The recombination time
is comparable to the Hubble time for the mean gas density at $z\sim 7$ and
so ionized regions persist \cite{Oh04} on these large scales where
inhomogeneities are small.  The minimum galaxy mass is increased by at least
an order of magnitude to a virial temperature of $\sim 10^5$K in these
ionized regions \cite{Bar01}.  It would be particularly interesting to
examine whether the faint end ($\sigma_\star < 30{\rm km~s^{-1}}$) of the
luminosity function of dwarf galaxies shows any moduluation on large-scales
around rare massive BHs, such as M87.

To find the volume filling fraction of relic regions from $z\sim 6$, we
consider a BH of mass $M_{\rm bh}\sim3\times10^9M_\odot$. We can estimate
the co-moving density of BHs directly from the observed quasar luminosity
function and our estimate of quasar lifetime.  At $z\sim 6$, quasars
powered by $M_{\rm bh}\sim3\times10^9M_\odot$ BHs had a comoving density of
$\sim 0.5\mbox{Gpc}^{-3}$\cite{WL03}.  However, the Hubble time exceeds
$t_{\rm dyn}$ by a factor of $\sim 2\times 10^2$ (reflecting the square
root of the overdensity in cores of galaxies), so that the comoving density
of the bubbles created by the $z\sim 6$ BHs is $\sim10^2\mbox{Gpc}^{-3}$
(see Fig. \ref{fig2}). The density implies that the volume filling fraction
of relic $z\sim 6$ regions is small, $<10\%$, and that the nearest BH that
had $M_{\rm bh}\sim3\times10^9M_\odot$ at $z\sim 6$ (and could have been
detected as an {\it SDSS} quasar then) should be at a distance $d_{\rm
bh}\sim \left(4\pi/3\times10^2\right)^{1/3}\mbox{Gpc} \sim140\mbox{Mpc}$
which is almost an order-of-magnitude larger than the distance of M87, a
galaxy known to possess a BH of this mass \cite{For94}.

{\it What is the most massive BH that can be detected dynamically in a
local galaxy redshift survey?} {\it SDSS} probes a volume of
$\sim1\mbox{Gpc}^3$ out to a distance $\sim30$ times that of M87.  At the
peak of quasar activity at $z\sim 3$, the density of the brightest quasars
implies that there should be $\sim100$ BHs with masses of
$3\times10^{10}M_\odot$ per $\mbox{Gpc}^{3}$, the nearest of which will be
at a distance $d_{\rm bh}\sim130\mbox{Mpc}$, or $\sim 7$ times the distance
to M87.  The radius of gravitational influence of the BH scales as $M_{\rm
bh}/v_{\rm c}^2\propto M_{\rm bh}^{3/5}$. We find that for the nearest
$3\times10^9M_\odot$ and $3\times10^{10}M_\odot$ BHs, the angular radius of
influence should be similar.  Thus, the dynamical signature of $\sim
3\times 10^{10}M_\odot$ BHs on their stellar host should be detectable.

\section{What seeded the growth of the supermassive black holes?}

The BHs powering the bright {\it SDSS} quasars possess a mass of a few
$\times 10^9 M_\odot$, and reside in galaxies with a velocity dispersion of
$\sim 500 {\rm km~s^{-1}}$\cite{Bark03}.  A quasar radiating at its
Eddington limiting luminosity, $L_E=1.4\times 10^{46}~{\rm
erg~s^{-1}}(M_{\rm bh}/10^8M_\odot)$, with a radiative efficiency,
$\epsilon_{\rm rad}=L_{E}/{\dot M}c^2$ would grow exponentially in mass as
a function of time $t$, $M_{\rm bh} =M_{\rm seed}\exp\{t/t_E\}$ on a time
scale, $t_E=4.1\times 10^7~{\rm yr} (\epsilon_{\rm rad}/0.1)$. Thus, the
required growth time in units of the Hubble time $t_{\rm hubble}= 9\times
10^8~{\rm yr}[(1+z)/7]^{-3/2}$ is
\begin{equation}
{t_{\rm growth}\over t_{\rm hubble}}=0.7 \left({\epsilon_{\rm rad} \over
10\%}\right) \left({1+z\over 7}\right)^{3/2}\ln \left({{M_{\rm
bh}/10^9M_\odot} \over M_{\rm seed}/100M_\odot}\right) ~.
\end{equation}
The age of the universe at $z\sim 6$ provides just sufficient time to grow
an {\it SDSS} BH with $M_{\rm bh}\sim 10^9M_\odot$ out of a stellar mass
seed with $\epsilon_{\rm rad}=10 \%$ \cite{Hai01}. The growth time is
shorter for smaller radiative efficiencies, as expected if the seed
originates from the optically-thick collapse of a supermassive star (in
which case $M_{\rm seed}$ in the logarithmic factor is also larger).

\begin{figure}[ht]
\begin{center}
\includegraphics[width=.8\textwidth]{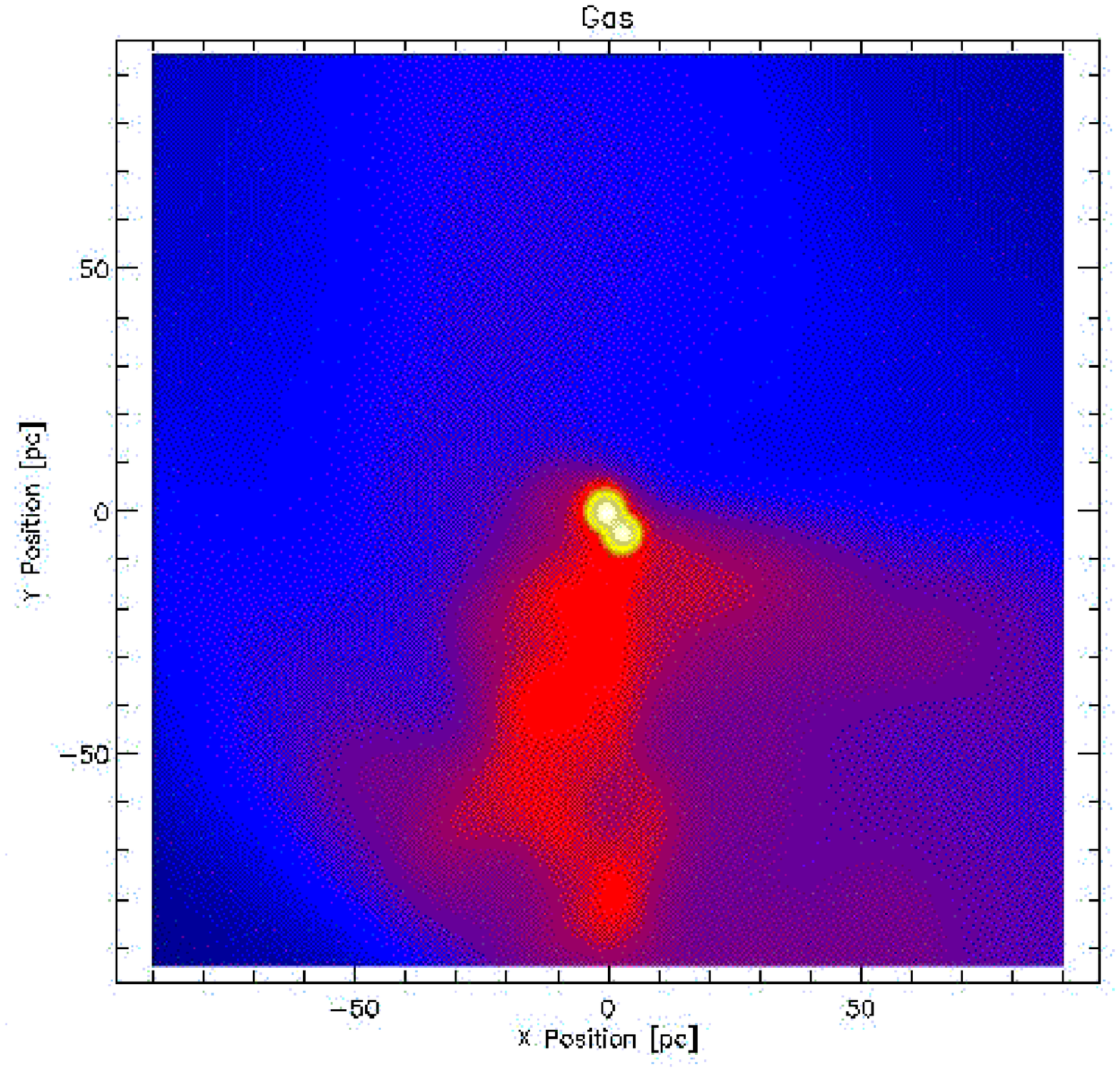}
\end{center}
\caption[]{SPH simulation of the collapse of an early dwarf galaxy with a
virial temperature just above the cooling threshold of atomic hydrogen and
no H$_2$ (from \cite{Bro03}).  The image shows a snapshot of the gas
density distribution at $z\approx 10$, indicating the formation of two
compact objects near the center of the galaxy with masses of $2.2\times
10^{6}M_{\odot}$ and $3.1\times 10^{6}M_{\odot}$, respectively, and radii
$<1$ pc. Sub-fragmentation into lower mass clumps is inhibited as long as
molecular hydrogen is dissociated by a background UV flux.  These
circumstances lead to the formation of supermassive stars
\cite{Loe94} that inevitably collapse and trigger the birth of supermassive
black holes \cite{Loe94,Bau02}.  The box size is 200 pc.  }
\label{fig6}
\end{figure}

{\it What was the mass of the initial BH seeds?  Were they planted in early
dwarf galaxies through the collapse of massive, metal free (Pop-III) stars
(leading to $M_{\rm seed}$ of hundreds of solar masses) or through the
collapse of even more massive, i.e. supermassive, stars \cite{Loe94} ?}
~Bromm \& Loeb \cite{Bro03} have shown through a hydrodynamical simulation
(see Fig. \ref{fig6}) that supermassive stars were likely to form in early
galaxies at $z\sim 10$ in which the virial temperature was close to the
cooling threshold of atomic hydrogen, $\sim 10^4$K. The gas in these
galaxies condensed into massive $\sim 10^6M_\odot$ clumps (the progenitors
of supermassive stars), rather than fragmenting into many small clumps (the
progenitors of stars), as it does in environments that are much hotter than
the cooling threshold. This formation channel requires that a galaxy be
close to its cooling threshold and immersed in a UV background that
dissociates molecular hydrogen in it. These requirements should make this
channel sufficiently rare, so as not to overproduce the cosmic mass density
of supermassive BH.

The minimum seed BH mass can be identified observationally through the
detection of gravitational waves from BH binaries with {\it Advanced LIGO}
\cite{WL04a} or with {\it LISA} \cite{WL03b}.  Most of the mHz binary
coalescence events originate at $z>7$ if the earliest galaxies included BHs
that obey the $M_{\rm bh}$--$v_c$ relation in Eq. (\ref{eq:1}). The number
of {\it LISA} sources per unit redshift per year should drop substantially
after reionization, when the minimum mass of galaxies increased due to
photo-ionization heating of the intergalactic medium.  Studies of the
highest redshift sources among the few hundred detectable events per year,
will provide unique information about the physics and history of BH growth
in galaxies \cite{Vol04}.

The early BH progenitors can also be detected as unresolved point sources,
using the future {\it James Webb Space Telescope} ({\it
JWST}). Unfortunately, the spectrum of metal-free massive and supermassive
stars is the same, since their surface temperature $\sim 10^5$K is
independent of mass \cite{Bro01}. Hence, an unresolved cluster of massive
early stars would show the same spectrum as a supermassive star of the same
total mass.

In closing, it is difficult to ignore the possible environmental impact of
quasars on {\it anthropic} selection. One may wonder whether it is not a
coincidence that our Milky-Way Galaxy has a relatively modest BH mass of
only a few million solar masses in that the energy output from a much more
massive (e.g. $\sim 10^9M_\odot$) black hole would have disrupted the
evolution of life on our planet. A proper calculation remains to be done
(as in the context of nearby Gamma-Ray Bursts \cite{Sca02}) in order to
demonstrate any such link.

\bigskip
\noindent
{\bf Acknowledgements.} I thank the collaborators who inspired my work on
this subject: Rennan Barkana, Volker Bromm, Steve Furlanetto, Zoltan
Haiman, and Stuart Wyithe.  This work was supported in part by NASA grant
NAG 5-13292, and by NSF grants AST-0071019, AST-0204514.

\end{document}